# Diffractive optics approach towards subwavelength pixels

Bo Fan[a], Sandeep Inampudi[a,b], Viktor A. Podolskiy*[a]

[a]Dept. of Physics and Applied Physics, Univ. of Massachusetts at Lowell, One University Ave., Lowell, MA 01854; [b](present address) Electrical and Computer Engineering Dept., Northeastern Univ., 360 Huntington Avenue, Boston, MA 02115

## ABSTRACT

Pixel size in cameras and other refractive imaging devices is typically limited by the free-space diffraction. However, a vast majority of semiconductor-based detectors are based on materials with substantially high refractive index. We demonstrate that diffractive optics can be used to take advantage of this high refractive index to reduce effective pixel size of the sensors below free-space diffraction limit. At the same time, diffractive systems encode both amplitude and phase information about the incoming beam into multiple pixels, offering the platform for noise-tolerant imaging with dynamical refocusing. We explore the opportunities opened by high index diffractive optics to reduce sensor size and increase signal-to-noise ratio of imaging structures.

## 1. INTRODUCTION

X-Ray, UV, optical, infrared, THz, and GHz radiation surround virtually every aspect of our life. Imaging is the process of recovering the parameters of the incoming radiation based on the analysis of its spatial-temporal field profile across a sensor[1,2]. In a sense, during imaging process, information carried by the incoming radiation is translated into recorded "image". In a simplest case when the operating frequency of the imaging/sensing system is fixed, the performance of the imaging system is limited by diffraction limit, which confines the linear size of individual pixel to approximately $\lambda_0/2$ (with $\lambda_0$ being free-space wavelength) or, equivalently, limits the angular resolution of the imaging system to approximately $\lambda_0/\Lambda$ (with $\Lambda$ being the linear size of the imaging system)[2]. The wavelength of light decreases in high-refractive-index materials according to $\lambda_{in} = \lambda_0/n$, potentially reducing pixel size and increasing angular resolution. In this work we analyze these phenomena in more details and discuss the benefits of using high-index media to improve imaging.

Although we restrict analytical and numerical parts of this work to the single-frequency planar imaging (see schematics in Fig.1), the approach presented below can be utilized in multi-frequency or white-light systems that operate in the linear optics regime. Furthermore, to clarify our main results, we restrict our discussions to systems that can directly characterize spatial profile of the electromagnetic field. At present, such systems are restricted to low (MHz…THz) operating frequencies [3]. Extension of our results to intensity-based systems will be presented in future work.

### 1.1 Diffraction limit and pixel size

Propagation of light in materials is governed by Maxwell equations. Linearity of Maxwell equations allows to represent a general solution as a linear combination of fundamental modes. For planar (or quasi-planar) layered systems plane waves, parameterized by their in-plane wavenumber $k_x$, represent a convenient choice of modal basis. Mathematically, distribution of electromagnetic fields of a plane electromagnetic wave in any given layer of the structure is proportional to $\exp(-i\omega t + i\vec{k} \cdot \vec{r})$ with components of the wavevector $\vec{k}$ related to angular frequency $\omega = 2\pi c/\lambda_0$ via

$$k_x^2 + k_y^2 + k_z^2 = \epsilon \frac{\omega^2}{c^2}$$

where $\epsilon = n^2$ represents relative permittivity of the material. For simplicity, we assume the layers to be isotropic and non-magnetic.

*viktor_podolskiy@uml.edu; phone 1 978 934-3398;

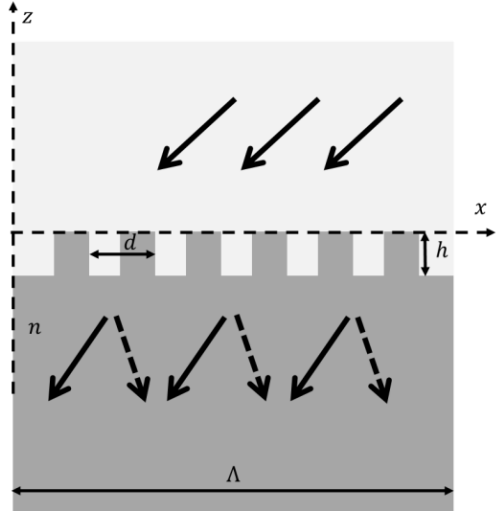

Fig.1 Schematics geometry of the diffractive imaging described in this work. Incident radiation enters high-refractive-index detector area through a diffraction grating. The detector "measures" the field along the line parallel to $x$ axis.

The spatial distribution of the incoming electromagnetic field is related to its spectral representation via

$$H_y(\vec{r}) = \int_{-n\frac{\omega}{c}}^{n\frac{\omega}{c}} A(k_x) \exp(-i\omega t + i \vec{k} \cdot \vec{r}) \, dk_x$$

Assuming in-plane propagation ($k_y = 0$), it is seen that the range of the propagating waves ($k_z^2 > 0$) is limited to $k_x \in \left[-n\frac{\omega}{c}, n\frac{\omega}{c}\right]$.

According to Fourier theory, the spatial width of the wavepacket represented by $H_y$ in the real space is inversely proportional to the spatial width of its spectrum in the wavenumber space. Therefore, the smallest size of such wavepacket (representing the "meaningful" smallest pixel size of the imaging system) is

$$\Delta_x \sim \frac{2\pi c}{2n\omega} = \frac{\lambda_0}{2n} = \frac{\lambda_{in}}{2}$$

This relationship is often referred to as *diffraction limit*[2,4]. Note that the analysis above is an order-of-magnitude estimate. The exact proportionality between the sizes of the wavepacket and its spectral representation is a function of the wavepacket shape[5], and thus depends, at least, on the configuration of the imaging system.

In refractive system (in the system where all relevant sizes are much larger than the wavelength) it is the *maximal* wavelength across the structure that determines the pixel size. For free-space-based imaging, the pixel size is limited by free-space wavelength. Diffraction limit prevents creation of static compact imagers operating at relatively low (IR…MHz) frequencies.

**1.2 Diffraction limit and the angular resolution**

The relationship between the size of the wavepacket and the spectral width has another interesting consequence. Just like the finite bandwidth of the wavepacket imposes the limitation on spatial separation between two neighboring pixels the finite size of any practical detector imposes limitation on the angular resolution of the system [6]. Mathematically, angular resolution can be characterized by the separation between two neighboring wavenumbers resolvable by the detector of the size $\Lambda$,

$$\Delta_k \sim \frac{2\pi}{\Lambda}$$

One way to arrive to this relationship is to assume that $N = 2\Lambda/\lambda_{in}$ pixels are placed on the regular rectangular grid and to consider the discretized version of reversible Fourier transform[7], yielding the $N$ spectral components separated by $\Delta_k = 2n\omega/Nc$ each.

When the electromagnetic radiation enters dielectric material with relatively large refractive index, the minimum pixel size is reduced, and the number of the "meaningful pixels" is increased. However, since the bandwidth of the propagating wavevector spectrum is also increased, the minimum angular resolution is unchanged.

On the other hand, in the imaging process the incoming radiation enters the high-index medium from vacuum. Therefore, the "extra bandwidth" does not necessarily contain the "extra information", and may provide redundancy for the system. The goal of this work is to analyze the benefits offered by the high-index media for imaging problem.

## 2. MAIN RESULTS

### 2.1 General Formalism

In this work we utilize full-wave solutions of Maxwell equations with commercial finite-element-method (FEM) solver[8]. The geometry of the system is illustrated in Fig.1. The finite region of space is surrounded by the perfectly matched layers. The plane waves are generated by the line source, and the resulting field distribution is analyzed inside the high-index "detector" region. The lateral dimension of the detector region is set to $\Lambda = 10\lambda_0$.

The imaging is performed in the three-step process. In the first step, we create a map that, given a wavevector of the plane wave incident on the system, produces the spatial distribution of the field along the sensor. In the second step, we consider a wavepacket [parameterized by the amplitude distribution $A(k_x)$ incident on the structure] and utilize the $A(k_x) \to H_y(x)$ map to calculate the total field distribution along the sensor, produced by such a wavepacket. In the final step of the process, the field distribution along with the original map can are used to recover the incident spectrum via least-square-fitting, similar to the process of Ref.[9], completing the imaging cycle.

Several test systems are used in this work to assess the benefits of high-index surrounding and of diffractive optics. The "baseline" is a non-diffractive all-vacuum system. In remaining three systems detector region is filled with transparent material with refractive index of 4 ($\epsilon = 16$), interfaced with surrounding vacuum via a diffraction grating with period $d$ and height $h$.

As discussed in the introduction, the finite-sized detector has a spectral resolution of $\Delta_k = 2\pi/\Lambda$. The diffractive grating is designed in such a way so that it spectrally spreads out the diffracted beams. Explicitly, we use

$$d = \frac{10}{11}\lambda_0$$

In all structures used in this work diffractive gratings have 50% duty cycle. Apart from the baseline structure we use three diffractive systems with thicknesses $h = 0$ (smooth surface), $h = \lambda_0/100$ (thin grating), and $h = \lambda_0$ (thick grating).

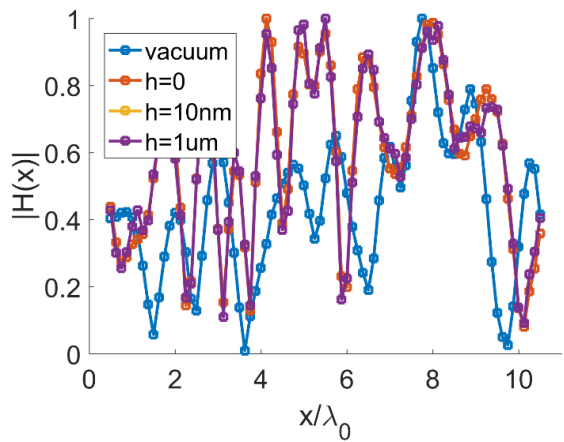

Fig.2 typical field distributions in three systems

Fig. 2 illustrates the typical field distributions along the detector in in each of the three systems. It is seen that the field distribution in the high-index structure oscillates faster than in its baseline counterpart, indicating the necessity of dense pixel arrays to detect these variations. As shown in Ref.[9], dense pixel arras can be used to perform holographic imaging based on field or intensity. Here we aim to quantify the limits of such diffractive imaging paradigm.

We note that the field distribution in the three diffractive structures is relatively similar to each other, indicating that the diffractive efficiency of our grating is relatively weak. Therefore, the results presented below should be used as typical estimates, not as best-case-scenario.

### 2.2 Spectral resolution of diffractive imaging systems

Spectral resolution of the imaging structures (the number of independent angular components characterized by distinct $k_x$ values) is related to the rank of the mapping matrix introduced above. However, in both numerical and experimental studies, the signal is always accompanied by some noise. Therefore, we characterize the rank of the matrix as the number of its eigenvalues larger than some cut-off value. The dependence of such rank on the eigenvalue cut-off tolerance for the four systems under consideration is outlined Fig.3. It is seen that when precise measurements are possible, diffractive optics provides significant (almost two-fold) improvement in angular resolution. However, in the limit of significant measurement noise the improvement in angular resolution is rather incremental.

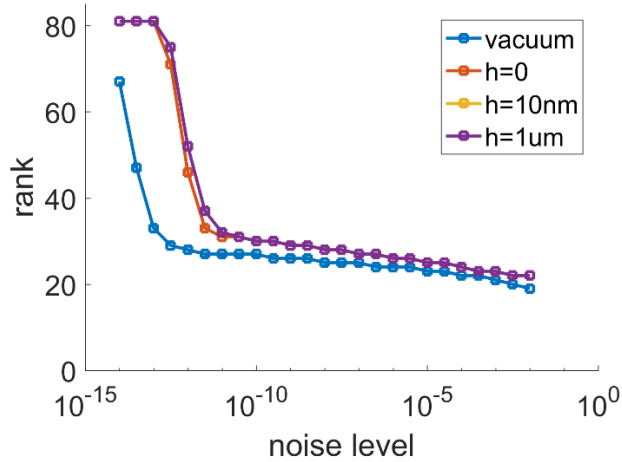

Fig.3 Rank of an $A(k_x) \rightarrow H_y(x)$ mapping function of the imaging system as a function of noise (cut-off eigenvalue amplitude)

Notably, the smooth high-index detector provides an improvement that is comparable to that of relatively thick diffractive grating. Our analysis of field distribution suggests that the benefits of diffractive structures can be further enhanced by optimizing their diffractive efficiency [10].

### 2.3 Noise tolerance of diffractive imaging

In the second part of this project we fixed the spectral density of the incoming radiation to the diffraction limit $\Delta_k$, as defined in the Introduction, and assumed that detector adds some noise to the measured field. Both amplitude and phase of the noise at every pixel are assumed to be randomly distributed functions.

To understand the effect of such a noise on the imaging performance of the system, we analyzed recovery of a set of incident field distributions. In each imaging attempt, FEM-generated map was used to calculate the "perfect" signal at the detector. The signal was then "degraded" by adding a random noise to each pixel. Finally, the map was when used to recover the original spectrum. The accuracy of the imaging process is given by the mean-square deviation between the spectral amplitudes of the incident light and recovered spectral amplitudes.

For each level of noise amplitude (defined as a fraction of maximum signal amplitude), we have analyzed recovery of 100 randomly generated incident field distributions. Results of this study are summarized in Fig. 4. The figure also compares the noise in high-index diffractive structures to the noise tolerance in vacuum-based diffractive structures that can be realized in X-Ray optics regime [11].

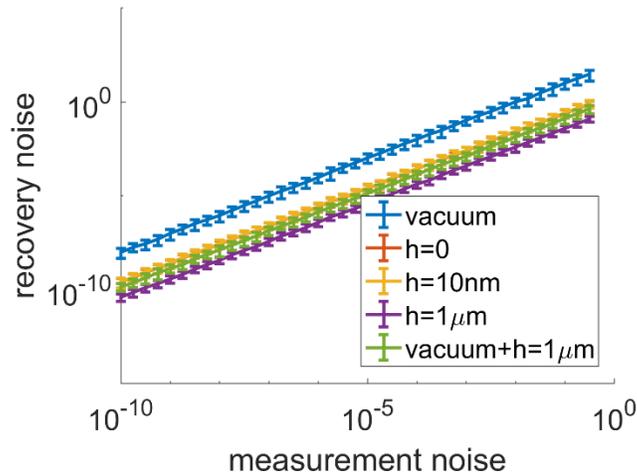

Fig.4. Dependence of recovery accuracy on the measurement noise amplitude

It is seen that diffractive optics provides significant improvement in noise-tolerance as compared with both baseline vacuum system as well as with smooth (and almost-smooth) interfaces.

## 3. CONCLUSION AND DISCUSSION

We expect that the results presented in this work can be reproduced with intensity-based holographic imaging and with multi-dimensional imaging structures. Further research is needed for detailed analysis of stability with respect to different numerical solvers of Maxwell equations. Specifically, to analyze effects related to the finite-size detector, our model used a finite simulation domain surrounded by perfectly matched layers. Finite linear source was used to excite the system. Therefore, the results of this study may be affected by diffraction at the edges of the source as well as by diffraction at the edges of the grating and, possibly, PML interfaces. Our previous studies [9] have indicated that extra noise could be added to the imaging process when different numerical techniques (such as FEM and Rigorous Coupled Wave Analysis) are used to generate the map and to generate the "source" distribution of the field, to be analyzed with the help of such a map. Here, FEM software was used for both purposes.

Our results also suggest that the angular resolution can be further enhanced by optimizing diffraction efficiency of the gratings and other diffractive structures, such as metasurfaces [12], for example, via a process similar to what has been reported in Ref.[13,14].

To conclude, we have analyzed spectral resolution and noise tolerance of the diffractive imaging paradigm, introduced in Ref.[9]. We have demonstrated that in diffractive imaging structures can potentially double angular resolution of the imaging systems in the limit of low measurement noise. In a separate study we demonstrated that diffractive structures can be used to significantly enhance noise tolerance of field-based imaging.

This work has been supported by US Army Research Office (Grant #W911NF-16-1-0261) and National Science Foundation (Grant #DMR-16293330)